\documentclass[12pt]{iopart}

\usepackage{graphicx,float}
\usepackage{longtable}
\begin{document}

\title[Submitted to Chinese Physics C]{Theoretical study on favored alpha-decay half-lives of
deformed nuclei
}

\author{M.Hassanzad \& O.N.Ghodsi}

\address{Department of Physics, Faculty of Basic Sciences, University of Mazandaran, P.O.Box 47416-416, Babolsar, Iran}
\ead{mo.hassanzad@gmail.com}
\vspace{10pt}
\begin{indented}
\item[]2021
\end{indented}

\begin{abstract}
A systematic study on $\alpha$-decay half-lives for the nuclei in the range $93\leq Z \leq 118$ is done by employing various versions of proximity potentials. To obtain the more reliable results, the deformation terms are included up to hexadecapole ($\beta_{4}$) in the spherical-deformed nuclear and Coulomb interaction potentials. The favored $\alpha$-decays process in this region are categorized as the even-even, odd A, and odd-odd nuclei, first, and second they are grouped into two parts as a ground state to ground state and ground state to isomeric states transition. Due to their root-mean-square deviations (RMSD's) comparison, $Bass 77$ and $Ngo 80$ have the least values and better compromise in reproducing experimental data. Besides, by considering preformation probability within the cluster formation model, the results validate that the significant reduction in root-mean-square deviations has been obtained for different versions of proximity; Hence detract the deviation between calculated and experimental data.
\end{abstract}

%\maketitle

\section{INTRODUCTION}
\label{sec:1}

To data transuranium nuclide with Z$\leq$118 have been synthesized, which has been experimentally and theoretically studied \cite{0010,0020,0030,0040,B1}. One of the most dominant decays for these radioactive nuclei is $\alpha$-decay, which is a powerful and precise tool to probe on the nuclear structure, including half-life, $\alpha$-clustering, shell effect, and deformation \cite{0050,0051,0060,0070,0080,0090,0100,0100}.

From the theoretical side, the $\alpha$-decay half-life depends on penetration probability and preformation factor, primarily. The former one has been calculated within the framework of Wntzel-Kramer-Brillouin (WKB) approximation which is mainly sensitive to the interaction potential between $\alpha$-daughter particles and also $\alpha$-particle energy. Concerning the quantum tunneling concept \cite{0110,0120}, a confined $\alpha$-particle, which carrying kinetic energy in the parent nucleus potential with non-zero penetration probability, has to escape from the potential barrier. Thereby, the choice of potential affects the half-life values. Hence, different theoretical methods have been suggested and developed such as generalized liquid drop model (GLDM) \cite{0131,0130,0140}, The density-dependent M3Y interaction and the mean-field potential \cite{0150,0151}, and empirical formulas \cite{0141,0142,0143,B2}.

When it comes to the latter one, the preformation factor ($P_{\alpha}$) refers to the probability of finding $\alpha$-cluster inside the parent nucleus. Determination of this factor can make available more nuclear structural information. Its probability magnitude should be less than or equal to one \cite{0210}. This quantity can be obtained from the ratios of calculation to experimental $\alpha$-decay half-life \cite{0240,0C1,0C2,0250}; so the theoretical model which produced different penetration probability has a crucial role in computing $P_{\alpha}$ in this way. Besides, the $P_{\alpha}$ values can be obtained from the overlap between the wave function of the parent nucleus and the coupled-wave function of the $\alpha$-daughter nucleus after the decay process \cite{0220,0230}.

Recently, cluster formation model (CFM) \cite{0260,0270,0280,0290,0300} was proposed to extract
the $\alpha$-preformation factor in terms of the $\alpha$-cluster formation energy based on the binding energy differences of the participating nuclide, which is in good agreement with different microscopical approaches. Deng \emph{et al.} \cite{0310} studied the $\alpha$-decay half-lives for nuclei around Z = 82, N = 126 closed shells by the proximity potential 1977 formalism; they confirmed that the effective and microscopic $P_{\alpha}$ within the CFM reduce the discrepancy between theoretical and experimental data in the mentioned region.

The proximity model is used to calculate the nuclear interaction energy between two nuclei by implying the surface energy. It is satisfactorily used in producing the nuclear potential between two nuclei while they are considered spherical \cite{0170,0180,0190,0200}. This theory has been derived, in a more complicated way, for the nuclear interaction between deformed–deformed pairs of nuclei\cite{A1, A2, A3}; Also, one of the interacting nuclei can be spherical. Due to this theorem, different physical quantities and phenomena have been studied \cite{A4, A5, A6, A7}.

Nuclei have many excited states that $\alpha$-decay can occur from the ground state or isomeric state of the parent nucleus to the ground state or isomeric state of the daughter nucleus; Here we consider those which decay from the ground state to the ground state or isomeric state with the same spin and parity which their minimum angular momentum transition is equal to zero, called favored $\alpha$-decays \cite{B3, B4}. The main objective of the current paper is to take the deformation of daughter nuclei in the favored $\alpha$-decay process under consideration. This article is organized as follows. The theoretical framework is summarized in Section \ref{sec:2}. The Calculations and results are discussed in Section \ref{sec:3}. And the conclusion of the entire work in Section \ref{sec:4}.

\section{THEORETICAL FRAMEWORK}
\label{sec:2}

$\alpha$-decay half-life of a parent nucleus can be determined as $T_{1/2}=\ln2/P_{\alpha}\nu_{0} P$. Here, $ P_{\alpha}$ is the preformation probabilities which will have discussed by next section and the assault frequency $\nu_0$ is related to the oscillation frequency $\omega$ \cite{0320}

\begin{equation}
\label{eq1}
{\nu_{0}}={\frac{\omega}{2\pi}}={\frac{(2n_{r}+l+\frac{3}{2})\hbar}{(2\pi\mu R_{n}^2)}}={\frac{(G+\frac{3}{2})\hbar}{(1.2\pi\mu R_{0}^2 )}},
\end{equation}
where $R_{n}^2=\frac{3}{5} R_{0}^2$  \cite{0330} and due to the under study in this paper, the global quantum number $G$ is equal to 22 \cite{0340}.

The $\alpha$-decay penetration probability, in a different orientation, using the WKB semi-classical approximation defined as

\begin{equation}
\label{eq2}
P=exp \left( {{-\frac{2}{\hbar}} \int^{r_{b}}_{r_{a}} \sqrt{2\mu(V_{T}-Q_{\alpha})}}dr \right),
\end{equation}
where ${\mu}= m(A_{\alpha}+A_{d})/A_{\alpha} A_{d}$ is the reduced mass which $A_{\alpha}=4$ and $A_{d}$ is daughter nucleus. The $r_{a}$ and $r_{b}$ are the turning points, which obtain from $V_{T}(r_{a} )=Q_{\alpha}=V_{T} (r_{b})$. Total interaction potential $V_{T}=V_{N}+V_{C}+V_{l}$ between $\alpha$-particle and daughter nucleus is defined as the sum of the Nuclear, Coulomb, and centrifugal potential, respectively, which have deformation and orientation dependence.

The nuclear term is introduced with detail in Ref \cite{0170} (and ref. therein) which is including different modifications on $Prox.77$ that are indicated in Table \ref{tab:1} and other versions of proximity potentials. For the details of the deformation effect \cite{A2, A3}, the mean curvature radius $\overline{R}$ with azimuthal angle $\phi$ between the principal planes of curvature of two deformed nuclei is given by

\begin{table}[h]
\caption{\label{tab:1} Prox 77 and its different modifications corresponding table 1 of Ref.\cite{0170}.}

\begin{indented}
\lineup
\item[]\begin{tabular}{@{}*{7}{l}}

\br
Proximity &            &            &Proximity &            &             \\
version   & $\gamma_0$ & $\kappa_s$ & version  & $\gamma_0$ & $\kappa_s$ \\
\mr
$Prox.66 $       &  1.01734   &  1.79    &  $Prox.81-III $   &  1.2502    &  2.4       \\
$Prox.76 $       &  1.460734  &  4       &  $Prox.88   $     &  1.2496    &  2.3       \\
$Prox.79 $       &  1.2402    &  3       &  $Prox.95   $     &  1.25284   &  2.345     \\
$Prox.81-I   $   &  1.1756    &  2.2     &  $Prox.03-I  $    &  1.08948   &  1.983     \\
$Prox.81-II  $   &  1.27326   &  2.5     &  $ModProx.88 $    &  1.65      &  2.3       \\
\br
\end{tabular}
\end{indented}
\end{table}

\begin{eqnarray}
\frac{1}{\overline{R^{2}}}= \frac{1}{R_{11}R_{12}} + \frac{1}{R_{21}R_{22}} + \left[\frac{1}{R_{11}R_{21}} + \frac{1}{R_{12}R_{22}}\right]sin^{2}\phi \nonumber\\
+ \left[\frac{1}{R_{11}R_{22}} + \frac{1}{R_{21}R_{12}}\right]cos^{2}\phi,\label{eq4}
\end{eqnarray}

where in this study nuclei are considered in the same plane so $\phi$ is equal to zero. $R_{i1}(\alpha_{i})$ and $R_{i2}(\alpha_{i})$ $(i=1,2)$ are the radii of curvature in the principal planes of each of the two nuclei

\begin{equation}
\label{eq5}
R_{i1}(\alpha_{i}) = \left|\frac{[R^{2}_{i}(\alpha_{i})+ R^{'2}_{i}(\alpha_{i})]^{3/2}}{R^{2}_{i}(\alpha_{i})+ 2R^{'2}_{i}(\alpha_{i})-R_{i}(\alpha_{i}) R^{''}_{i}(\alpha_{i})}\right|
\end{equation}

\begin{equation}
\label{eq6}
R_{i2}(\alpha_{i}) = \left| \frac{R_{i}(\alpha_{i})sin(\alpha_{i})+ [R^{2}_{i}(\alpha_{i})+R^{'2}_{i}(\alpha_{i})]^{3/2}} {R^{'}_{i}(\alpha_{i})cos(\alpha_{i})+R^{'}_{i}(\alpha_{i})sin(\alpha_{i})} \right|
\end{equation}

with separation distance, R, between their centers their minimum distance is defined as

\begin{equation}
\label{eq7}
s= \left| R-R_{1}(\alpha_{1})-R_{2}(\alpha_{2}) \right|  ,
\end{equation}

with

\begin{equation}
\label{eq8}
{r_i (\alpha_i)}= {r_{0i} \left[1 + \sum_{\lambda} \beta_{\lambda i} Y_{\lambda}^{(0)} (\alpha_i) \right] },
\end{equation}

where $r_{0i}=1.28 A_{i}^{1/3}-0.76+0.8 A_{i}^{-1/3}$. Here $\alpha_i$ is the angle between the radius vector and the symmetry axis of the $i^{th}$ nuclei. In this study we consider one of two nuclei is spherical, So, it has no deformation parameters.

The rotational effect of two nucleus systems can be calculated by the l-dependent centrifugal potential which is equal to $\hbar^{2} l(l+1)/2\mu r^{2}$. The $l$ is the orbital angular momentum carried by the $\alpha$-particle, these values that are used to calculate this potential are determined by using the permitted transitions between the parent nucleus and the daughter nucleus.

The Coulomb potential between spherical-deformed and oriented, taken from Ref.\cite{0360} is given as follows

\begin{equation}
\label{9}
V_{C}(r,\beta_{\lambda},\theta) = \cases{ Z_{\alpha} Z_{d} e^2 \left\{\frac{1}{r} + \frac{3}{2 \lambda +1} \frac{r_{T} ^{\lambda}}{r^{\lambda +1}} \beta_{\lambda} Y_{\lambda,0}(\theta)\right\}   &  for $r \ge r_{01}$\\
Z_{\alpha} Z_{d} e^2 \left\{\frac{1}{2r_{T}}\left[3-(\frac{r}{r_{T}})^2\right]+\frac{3}{2\lambda+1}\frac{r^{\lambda}}{r_{T}^{\lambda +1}}\beta_{\lambda}Y_{\lambda,0}(\theta)\right\}    &  for $r \leq r_{01}$\\}
\end{equation}

This potential has both good accuracy and time-saving in calculation. Because the deformation parameters from M\"{o}ller $\emph{et al}$  \cite{0370} agree well with the existent magnitude of experimental deformation \cite{0380}, $\beta_{\lambda}$ are taken from \cite{0370} for all calculations of $\alpha$-decay half-lives of deformed nuclei in this study.

\section{RESULTS AND DISCUSSION}
\label{sec:3}

\begin{figure}[H]
\includegraphics [width=1.0\textwidth,origin=c,angle=0] {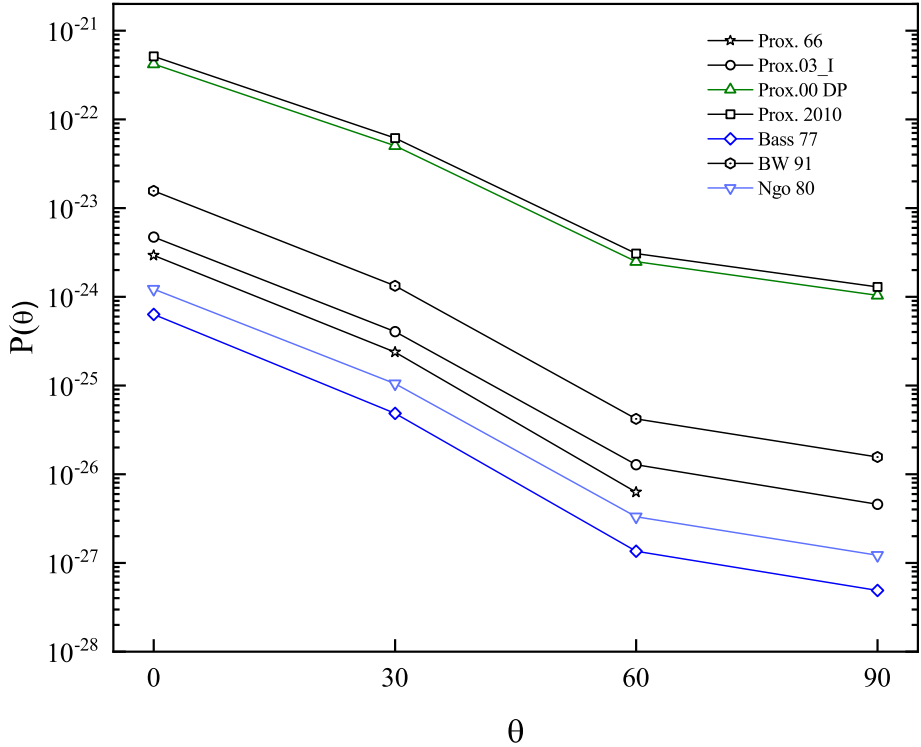} % Here is how to import EPS art
\caption{\label{fig:figure1} The penetration probability of the eight selected versions of proximity potentials with respect to the angles of emitted $\alpha$-particle.}
\end{figure}

We have studied 108 transuranium nuclei in the wide atomic range of $Z=93$ onward, Systematically, by using 23 versions of spherical-deformed potential for both the short-range attractive term and the Coulomb repulsive term of the potential barrier, in which the $\alpha$-particle and the residual nuclei are respectively considered spherical and deformed. It is interesting to see how the deformation affects the Proximity potentials and the corresponding penetration probability in the $\alpha$-decay process. For instance, the penetration probability for $^{253}$Fm $\rightarrow$ $^{249}$Cf+$\alpha$ (with $\beta_{2}=0.250$ and $\beta_{4}=0.039$) is plotted in each direction in Fig.\ref{fig:figure1} for seven selected versions. Corresponding to this figure, the nuclear potential for decay through $\theta=0^{\circ}$ is deeper than $\theta=90^{\circ}$. Due to the proximity theory \emph{(The force between two gently curved surfaces as a function of the separation degree of freedoms is proportional to the interaction potential per unit area)}, one can realize that a nucleus contains a thicker surface at $\theta=0^{\circ}$. We can figure out that the stronger interaction in this area is due to the overlap of the nucleons. Although properties like the Pauli exclusion principle, spin and parity, isospin asymmetry, and so on play role in the formation of a particle before emitting, Nevertheless, One can implicitly be expected that the probability of formation of an $\alpha$-particle in this area is more likely than others. Also From this figure, we can explore that the $Q_{\alpha}$ line does not cross the total potential curve in some direction like prox. 66; provided that it arises in all directions, we are not able to calculate the penetration probability integral.

The height and position of the barrier potential of the ${\alpha}$-particle in a deformed nucleus depend on the angle that ${\alpha}$-particle is emitted. The average of penetrability is obtained by using

\begin{equation}
\label{eq9}
P(\theta)={\frac{1}{2} \int_{0}^{\pi} P \sin(\theta)d\theta }.
\end{equation}

In the case of the probability of the $\alpha$-particle formation before penetration, the cluster-formation model is an energy-dependent theory proposed to calculate the ${\alpha}$ preformation factor. The basic assumption in the CFM formalism is that the nucleons around the surface contribute to the preformation of the ${\alpha}$-particle. In this model, the ${\alpha}$-preformation factor is defined as $P_{\alpha}={E_{f\alpha} / E}$, where $E_{f\alpha}$ is the formation energy of the ${\alpha}$-cluster and $E$ is the total energy of a considered system \cite{0260,0270,0280,0290,0300}. In order to use the accurate mass excess which is vitally important to obtain the precise binding energies, the data are given from \cite{0390}.

We have computed $\alpha$-decay half-lives and evaluated a quantitative analysis of them by using root-mean-square deviation as $RMSD=\sqrt{1/N \sum_{i=1}^N [Log_{10}(T_{1/2,i}^{Exp.})-Log_{10}(T_{1/2,i}^{Theo.})]^2}$, where N is the number of contributed nuclei, and experimental values of half-lives are taken from \cite{0450} with takeing their branching ratios into account.

Since the parent nucleus can decay from ground state(gs) to different possible states that satisfy the spin-parity rule:
\begin{equation}
\label{eq9}
{|I_{i}-I_{j}| \leq l \leq I_{i}+I_{j}};~~~~~~~~~{\pi_{i}\pi_{j}=(-1)^{l}}.
\end{equation}

\begin{figure}[H]
\includegraphics [width=1.0\textwidth,origin=c,angle=0] {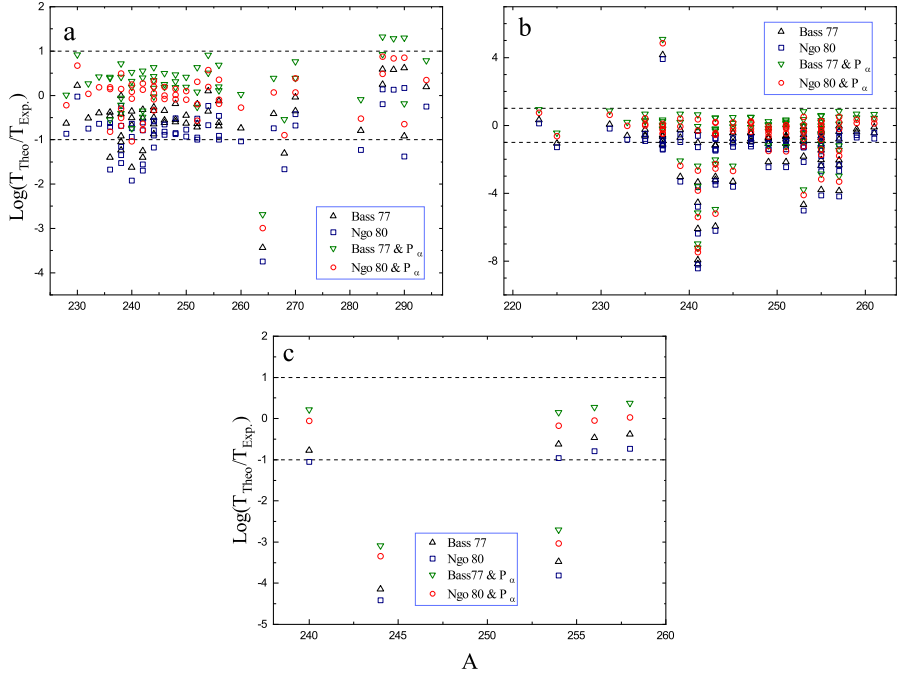} % Here is how to import EPS art
\caption{\label{fig:figure2} Deviation between calculated and experimental ${\alpha}$-decay half-lives for $a)$ even-even $b)$ odd A $c)$ odd-odd nuclei}
\end{figure}

According to this equation, different values of $l$ can transfer from the initial state to the final states, although, the minimum possible value of angular momentum is of our interest. Due to this value, the favored transition will occur provided $l=0$, else transition to the daughter nucleus is hindered. All considered nuclei in this study are favored ones, 22 nuclei are decay to more than one level with the same spin and parity. These total 141 favored $\alpha$-decays contains 54 even-even(e-e) nuclei, 81 odd A, and 6 odd-odd(o-o) nuclei. The experimental data and their details about the isomeric states and energy levels are taken from \cite{0450}. RMSDs of these groups are tabulated in table \ref{tab:2}, column two to four; their corresponding value after taking the preformation factor under consideration is given in columns five to seven. Comparing the RMSDs of different proximity versions for e-e nuclei shows that $Bass 77$ and $Ngo 80$ have the least value respectively. It is interesting that when P$_{\alpha}$ take under consideration, $Ngo 80$ obtains better results than $Bass 77$. The deviation between calculated and experimental values for e-e elements are shown in figure \ref{fig:figure2} a. One can find out that these versions reproduce the experimental values very well.

\begin{longtable} [h]{lllllll}
\caption{\label{tab:2} RMSDs of ${\alpha}-$decay half lives of even-even, odd A,and odd-odd nuclei considering $P_{\alpha}$}\\

\hline\noalign{\smallskip}

           &   \centre{3}{$P_{\alpha}^{1}$}       &  \centre{3}{$P_{\alpha}^{CFM}$}       \\
\ns
Proximity  &       \crule{3}                        &     \crule{3}                         \\
Potentials & e-e & odd A  & o-o & e-e & odd A  & o-o \\

\hline\noalign{\smallskip}
\endfirsthead
\multicolumn{7}{l}%
{\tablename\ \thetable\ -- \textit{Continued from previous page}} \\
\hline\noalign{\smallskip}

           &   \centre{2}{$Q_{\alpha}^{1}$}     &         &  \centre{2}{$Q_{\alpha}^{CFM}$}      &          \\
\ns
Proximity  &       \crule{3}                      &     \crule{3}                          \\
Potentials & e-e & odd A  & o-o & e-e & odd A  & o-o \\

\hline\noalign{\smallskip}
\endhead
\hline\noalign{\smallskip} \multicolumn{7}{r}{\textit{Continued on next page}} \\
\endfoot
\hline\noalign{\smallskip}
\endlastfoot
$Prox 66 $    &  1.3660  &  2.6930   &  2.7633   &  0.7961   &  2.0759    & 2.0655 \\
$Prox.76 $    &  2.0879  &  3.3049   &  3.4166   &  1.3441   &  2.5680    & 2.6440 \\
$Prox.79 $    &  1.8156  &  3.0700   &  3.1693   &  1.0950   &  2.3686    & 2.4177 \\
$Prox81-I $   &  1.7723  &  3.0382   &  3.1384   &  1.0559   &  2.3430    & 2.3899 \\
$Prox81-II $  &  1.9395  &  3.1830   &  3.2929   &  1.2057   &  2.4640    & 2.5300 \\
$Prox81-III $ &  1.9064  &  3.1546   &  3.2624   &  1.1756   &  2.4401    & 2.5022 \\
$Prox.88 $    &  1.9165  &  3.1641   &  3.2737   &  1.1846   &  2.4482    & 2.5124 \\
$Prox.95 $    &  1.9177  &  3.1650   &  3.2741   &  1.1857   &  2.4488    & 2.5129 \\
$Prox03_I $   &  1.5815  &  2.8749   &  2.9624   &  0.8957   &  2.2122    & 2.2341 \\
$ModProx88 $  &  2.4822  &  3.6703   &  3.8121   &  1.7189   &  2.8951    & 3.0165 \\
$Prox.00 $    &  1.8242  &  3.0916   &  3.1951   &  1.1032   &  2.3868    & 2.4410 \\
$Prox.00DP $  &  3.5506  &  4.6520   &  4.8360   &  2.7690   &  3.8184    & 4.0094 \\
$Prox.2010 $  &  3.6436  &  4.7335   &  4.9148   &  2.8613   &  3.8966    & 4.0867 \\
$Bass1977 $   &  0.8281  &  2.2505   &  2.2620   &  0.6547   &  1.8153    & 1.6867 \\
$Bass1980 $   &  2.0303  &  3.2768   &  3.4024   &  1.2881   &  2.5461    & 2.6317 \\
$CW1976 $     &  2.8746  &  4.0475   &  4.2094   &  2.1009   &  3.2450    & 3.3981 \\
$BW1991 $     &  2.0596  &  3.3127   &  3.4370   &  1.3153   &  2.5778    & 2.6639 \\
$Ngo1980 $    &  1.0474  &  2.4430   &  2.4907   &  0.5963   &  1.9144    & 1.8443 \\
$Denisov $    &  4.3656  &  5.4038   &  5.6299   &  3.5799   &  4.5480    & 4.7921 \\
$Denisov DP$  &  5.4896  &  6.4895   &  6.7427   &  4.7003   &  5.6155    & 5.8963 \\
$AW95 $       &  1.8899  &  3.0318   &  3.1703   &  1.1799   &  2.3319    & 2.4164 \\
$Dutt2011 $   &  1.9200  &  3.1811   &  3.2933   &  1.1870   &  2.4637    & 2.5308 \\
$Guo2013 $    &  2.4293  &  3.6335   &  3.7766   &  1.6674   &  2.8625    & 2.9820 \\
\end{longtable}

For other two groups in table \ref{tab:2}, $Bass 77$ and $Ngo 80$ are the most suitable versions; which the deviation of the calculate values are represented in figure \ref{fig:figure2} b and c for odd A and o-o nuclei, respectively. $Prox.66$ is not able to reproduce the half-life of just $^{290}$Fl. It is useful to notice that the experimental data can also impact on the results; for instance, the $^{264}$Hs which is recognizable in figure \ref{fig:figure2} a the only A=264, its experimental $\alpha$-decay half-life is reported $1$s in NUBASE2020 \cite{0450}, while it is $1.08$ms in NUBASE2016 \cite{0470}, while its $\alpha$-decay half-life is $0.365$ms and $0.178$ms and after using P$_{\alpha}$ it becomes $2.07$ms and $1.01$ms respectively for $Bass 77$ and $Ngo 80$.

\begin{figure}[H]
\includegraphics [width=1.0\textwidth,origin=c,angle=0] {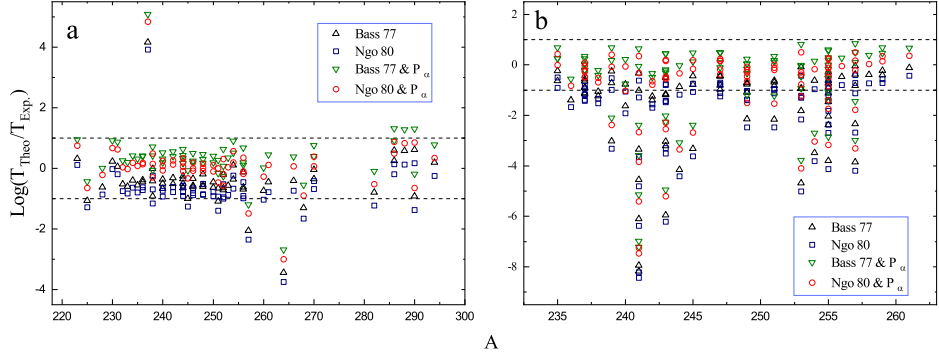} % Here is how to import EPS art
\caption{\label{fig:figure3} Deviation between calculated and experimental ${\alpha}$-decay half-lives for $a)$ from ground state to ground state and $b)$ from ground state to isomeric states.}
\end{figure}

To investigate the validity of these formalisms to estimate the ground state (gs) to the ground state or isomeric states (is), these 108 nuclei are grouped with respect to the energy level of the daughter nucleus that whether they are in their ground state or its isomeric states. Among the total 141 favored decays, 59 transitions are gs-gs and 82 transitions are gs-is, which their RMSD values are given in table \ref{tab:3}. Similar to table \ref{tab:2} their corresponding values after preformation values take under consideration are shown. For gs-gs nuclei, $Bass 77$ before and after using P$_{\alpha}$ is almost equal, whereas $Ngo 80$ with P$_{\alpha}$ is the outstanding versions. Figure \ref{fig:figure3} a, show the power of formalisms to produce experimental data of gs-gs nuclei. Furthermore, also $Bass 77$ and $Ngo 80$ can satisfy the gs-is elements better out of all. Figure \ref{fig:figure3} b shows the deviation between these suitable versions and experimental values.

\begin{longtable} [h]{lllll}
\caption{\label{tab:3} RMSDs of ${\alpha}-$decay half lives of ground state to ground state and isomeric states of nuclei are given considering $P_{\alpha}$.}\\

\hline\noalign{\smallskip}

           &   \centre{2}{$P_{\alpha}^{1}$}            &  \centre{2}{$P_{\alpha}^{CFM}$}              \\
\ns
Proximity  &       \crule{2}                         &     \crule{2}                          \\
Potentials & g.s.-g.s. &  g.s.-i.s.  & g.s.-g.s. & g.s.-i.s.  \\

\hline\noalign{\smallskip}
\endfirsthead
\multicolumn{5}{l}%
{\tablename\ \thetable\ -- \textit{Continued from previous page}} \\
\hline\noalign{\smallskip}

           &   \centre{2}{$Q_{\alpha}^{Exp.}$}      &  \centre{2}{$Q_{\alpha}^{WS4+}$}        \\
\ns
Proximity  &       \crule{2}                        &     \crule{2}                          \\
Potentials & g.s.-g.s. &  g.s.-i.s.  & g.s.-g.s. & g.s.-i.s.  \\

\hline\noalign{\smallskip}
\endhead
\hline\noalign{\smallskip} \multicolumn{5}{r}{\textit{Continued on next page}} \\
\endfoot
\hline\noalign{\smallskip}
\endlastfoot
$Prox 66	$ &    1.3865  &    2.7512  &  0.9531  &  2.0824 \\
$Prox.76	$ &    2.0324  &	3.3974  &  1.3535  &  2.6262 \\
$Prox.79	$ &    1.7798  &	3.1522  &  1.1483  &  2.4105 \\
$Prox81-I   $ &    1.7383  &	3.1197  &  1.1165  &  2.3828 \\
$Prox81-II  $ &    1.8912  &	3.2718  &  1.2351  &  2.5151 \\
$Prox81-III $ &	   1.8608  &	3.2421	&  1.2107  &  2.4890 \\
$Prox.88    $ &    1.8696  &	3.2522  &  1.2177  &  2.4979 \\
$Prox.95    $ &    1.8708  &	3.2531  &  1.2185  &  2.4987 \\
$Prox03_I   $ &    1.5683  &	2.9469  &  1.0010  &  2.2375 \\
$ModProx88  $ &	   2.3980  &	3.7802	&  1.6789  &  2.9771 \\
$Prox.00	$ &    1.7879  &	3.1749  &  1.1545  &  2.4302 \\
$Prox.00DP  $ &	   3.4408  &	4.7824	&  2.6785  &  3.9327 \\
$Prox.2010  $ &	   3.5346  &	4.8635	&  2.7704  &  4.0112 \\
$Bass1977   $ &    0.9399  &	2.2789  &  0.9338  &  1.7658 \\
$Bass1980   $ &    1.9770  &	3.3699  &  1.3058  &  2.6031 \\
$CW1976	    $ &    2.7592  &	4.1765  &  2.0167  &  3.3506 \\
$BW1991	    $ &    1.9986  &	3.4094  &  1.3222  &  2.6390 \\
$Ngo1980	$ &    1.1017  &	2.4897  &  0.8530  &  1.8920 \\
$Denisov	$ &    4.2357  &	5.5503  &  3.4625  &  4.6825 \\
$Denisov DP $ &	   5.3404  &	6.6522	&  4.5584  &  5.7699 \\
$AW95	    $ &    1.9032  &	3.0921  &  1.2804  &  2.3529 \\
$Dutt2011   $ &    1.8701  &	3.2712  &  1.2166  &  2.5154 \\
$Guo2013    $ &    2.3325  &	3.7493  &  1.6175  &  2.9484 \\
\end{longtable}

For the purpose of comparing our calculated half-lives with ones obtained from the analytical formula the recent form of universal decay law (UDL), \cite{0460} is used. The UDL for ${\alpha}$ and cluster decay modes was introduced as

\begin{equation}
\label{eq18}
Log_{10}(T_{\frac{1}{2}})=aZ_{c} Z_{d} \sqrt{\frac{A}{Q_{c}}} +b \sqrt{AZ_{c} Z_{d} \left(A_{d}^\frac{1}{3} + A_{c}^\frac{1}{3}\right)}+c,
\end{equation}

where $A=A_{c} A_{d}/(A_{c}+A_{d})$ and the constant $a=0.4314$, $b=-0.4087$ and $c=-25.7725$ are determined by fitting to experimental of both $\alpha$ and cluster decays. The RMSD values for UDL is equal to 0.6769, 1.8092,1.7120, 0.9650,and 1.7550 with respect to foe e-e, odd A, o-o, gs-gs, and gd-is, respectively.

It is noticeable that, the more the probability of formation $\alpha$-particle inside the parent's nucleus, the more the nucleus is unstable. Consequently, it is comprehensible that because form the $\alpha$-particles are more difficult around magic numbers, they are more stable and this issue is implied in computation. Furthermore, the preformation factor is made the half-life values bigger, so almost all versions considerably reduce the RMSDs, eventually, which is interpreted more proper to reproduce the half-lives.

%\newpage

\section{CONCLUSION}
\label{sec:4}

The wide range of favored $\alpha$-decay nuclei in atomic range 93$\leq$Z$\leq$118 has been taken under study using various versions of proximity potentials within the WKB approximation formalism to calculate their half-lives. In this study, for reproducing the nuclear and Coulomb barrier potential we consider the shape of the $\alpha$-particle and daughter nucleus as spherical and deformed, respectively. It is indicated that the ${\alpha}$-decay process is not an isotropic occurrence in space. Moreover, we have employed the CFM theory to estimate the preformation factors. On the one hand, we have analyzed these nuclei with respect to their number of proton and neutron, so they grouped into even-even, odd A, and odd-odd nuclei. On the other hand, considering the energy level of the daughter nucleus, they categorized as a ground state to ground state and ground state to isomeric states. The obtained results are indicated that one can choose $Bass 77$ and $Ngo 80$ as the most suitable versions among others to estimate the $\alpha$-decay half-lives; Also revealed that $\alpha$-preformation factor plays an invaluable role in half-life computation and decrease the RMSDs of all versions. These results can motive one's interest to develop and extend further research in the future.\\

\end{document}